\documentclass[reprint,amsmath,amssymb,aps,superscriptaddress,twocolumn,float]{revtex4-1}
\pdfoutput=1
\usepackage{graphicx}
\usepackage{graphics}
\usepackage{dcolumn}
\usepackage{bm}
\usepackage{amsfonts}
\usepackage{amssymb}
\usepackage{float}
\usepackage{xcolor}
\begin{document}
\title{Tunable large spin Hall and spin Nernst effects in Dirac semimetals \\
ZrXY (X=Si, Ge; Y=S, Se, Te)} 


\author{Yun Yen}
\affiliation{Department of Physics and Center for Theoretical Physics, National Taiwan University, Taipei 10617, Taiwan\looseness=-1}

\author{Guang-Yu Guo}
\email{gyguo@phys.ntu.edu.tw}
\affiliation{Department of Physics and Center for Theoretical Physics, National Taiwan University, Taipei 10617, Taiwan\looseness=-1}
\affiliation{Physics Division, National Center for Theoretical Sciences, Hsinchu 30013, Taiwan\looseness=-1}

\begin{abstract}

The ZrSiS-type compounds are Dirac semimetals and thus have been attracting considerable interest 
in recent years due to their topological electronic properties and possible technological applications. 
In particular, gapped Dirac nodes can possess large spin Berry curvatures and thus
give rise to large spin Hall effect (SHE) and spin Nernst effect (SNE), which may be used to
generate pure spin current for spintronics and spin caloritronics without applied magnetic
field or magnetic material. In this paper we study both SHE and SNE in ZrXY (X = Si, Ge; Y = S, Se, Te)
based on {\it ab initio} relativistic band structure calculations. 
Our theoretical calculations reveal that some of these compounds exhibit 
large intrinsic spin Hall conductivity (SHC) and spin Nernst conductivity (SNC). 
For example, the calculated SHC of ZrSiTe is as large as -755 ($\hbar$/e)(S/cm). 
Since the electric conductivity of these Dirac semimetals
are much smaller than that of platinum which has the largest intrinsic SHC 
of $\sim$2200 ($\hbar$/e)(S/cm), this indicates that they will have a larger spin Hall angle 
(and hence a more efficient charge-spin current conversion) than that of platinum.  
Remarkably, we find that both the magnitude and sign of the SHE and SNE in these compounds can be significantly 
tuned by changing either the electric field direction or spin current direction and
may also be optimized by slightly varying the Fermi level via chemical doping.
Analysis of the calculated band- and $k$-resolved spin Berry curvatures show that
the large SHE and SNE as well as their remarkable tunabilities originate from
the presence of many slightly spin-orbit coupling-gapped Dirac nodal lines
near the Fermi level in these Dirac semimetals.  
Our findings thus suggest that the ZrSiS-type compounds are promising candidates 
for spintronic and spin caloritronic devices, and will certainly stimulate
further experiments on these Dirac semimetals.

\end{abstract}

\maketitle




\section{INTRODUCTION}

Spintronics, a field combining the spin degree of freedom with traditional electronics, 
has drawn increasing attention over the past decades. The generation, detection and manipulation 
of spin current are the most important issues in spintronics. 
Several methods have already been developed to generate spin current. For example, spin current injection 
from ferromagnets to nonmagnetic materials, spin pumping \cite{switkes1999adiabatic}, 
and the method using spin momentum-locking surface states of topological 
materials\cite{hasan2010colloquium,liu2015spin}, were demonstrated. 
The spin Hall effect (SHE), first predicted in 1971 \cite{d1971spin}, is an unique spin 
current generation method without the need of applied magnetic field or a magnetic material. 
In a nonmagnetic sample under an applied electric field, because of the presence of spin-orbit coupling (SOC), 
electrons with opposite spins moving in the relativistic band structure would acquire 
opposite transverse velocities, and this results in a pure transverse spin current. 
The SHE has thus been extensively studied in recent years 
(see, e.g., \cite{hirsch1999spin,Mura03,Sinova04,Kato04,Guo05,Valenzuela06,Hoffman13,sinova2015spin} and references therein). 

Similarly, a longitudinal temperature gradient ($\nabla T$) could play the role 
of the applied electric field ($E$) in driving the electrons through the nonmagnetic sample, 
also leading to a pure transverse spin current. This thermally driven 
spin transport phenomenon in nonmagnetic materials is called the spin Nernst effect (SNE) \cite{sne2008}. 
It has already been experimentally observed in several materials such as Pt thin film \cite{meyer2017observation} 
and tungsten \cite{sheng2017spin}. This would make the spintronic devices powered by 
heat (known as spin caloritronics) possible. 
One can expect that the materials with large SHE could also possess large SNE, 
since within the Berry phase formalism they all come from the spin Berry curvatures 
in the relativistic band structure.~\cite{guo2008intrinsic,xiao2010berry,guo2017large}

The recent studies of SHE in topological insulators (TIs) such as Bi$_2$Se$_3$ \cite{mellnik2014spin}, 
have demonstrated that topological materials could have an efficient charge-spin current conversion. 
More recently, the search for the materials for spintronics has turned 
to three dimensional topological semimetals (TSMs). In the bulk band structure of the TSMs, 
stable nodal points and nodal lines exist~\cite{burkov2011topological} and their Fermi surfaces 
with a nontrivial topology would produce several interesting effects such as the Fermi arc
on their surfaces\cite{armitage2018weyl}. As a typical category of these TSMs, the Dirac semimetals (DSMs) 
are three dimensional analogues to graphene, 
with linearly dispersed excitation described by the Dirac Hamiltonian \cite{burkov2011topological}. 
Unlike the Dirac-like surface states in TIs, some robust Dirac points in DSMs are protected 
by crystalline symmetry \cite{young2015dirac,schoop2016dirac}, while unprotected ones would 
be gapped by the SOC. Just like the quantum spin Hall effect in graphene with the Dirac points gapped 
slightly by the SOC \cite{kane2005quantum,kane2005z}, 
the SHE in these DSMs could be also intriguing. For example, the recent studies on the SHE in the
rutile oxide Dirac semimetals \cite{sun2017dirac} and the TaAs Weyl semimetal family \cite{sun2016strong} 
showed that the anticrossing of the nodal points and 
nodal lines in these TSMs could be the sources of large spin Berry curvatures. 
 
Among the Dirac semimetals, the studies of the ZrSiS-type compounds have been particularly intense 
in the past few years. The members of the ZrSiS family share the same space group \textit{P}4/\textit{nmm} (No.129), 
and they all possess the symmetry protected Dirac nodal features in the band 
structures \cite{schoop2016dirac,hu2016evidence,hosen2018observation,Su2018}. In addition, several exotic quantum phenomena 
such as highly anisotropic magnetoresistance  \cite{ali2016butterfly}, gapless Dirac surface states 
in topological crystalline insulator (TCI) phase \cite{hosen2018observation}, Shubnikov-de Haas and de Haas-van Alphen 
oscillations \cite{hu2017quantum,Su2018}, have been observed in some compounds of the family. 
Nonetheless, the spin Hall effect and spin Nernst effect in such systems have not been studied yet. 

In this paper, therefore, we present the main results of our systematic {\it ab initio} density functional theory 
(DFT) calculations on the SHE and SNE in DSM compounds ZrSiS, ZrSiSe, ZrSiTe, ZrGeS, ZrGeSe and ZrGeTe. 
The rest of this paper is organized as follows.
In the next section, we briefly describe the Berry phase formalism for
calculating the intrinsic spin Hall and Nernst conductivities as well as the computational details.
Section III consists of four subsections.
We first present the calculated relativistic band structures of the six considered compounds
and also analyze their topological properties in Sec. III A.
We then report the calculated spin Hall conductivities, which are compared with that of some well-known
spin Hall metals, in Sec. III B. To understand the origins of the large SHC, 
we present the band- and $k$-resolved spin Berry curvatures especially near the
Dirac nodes in the vicinity of the Fermi level in Sec. III C.
We finally present the calculated spin Nernst conductivities and their temperature dependences,
and also discuss the validity of the Mott relation
in Sec. III D. Finally, the conclusions drawn from this work are summarized in Sec. IV.

\begin{figure}[t] \centering \includegraphics[width=8.5cm]{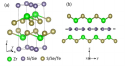}
\caption{Nonsymmorphic structure of the ZrSiS family. (a) Crystal structure, and (b) illustration of 
its nonsymmorphic glide mirror symmetry, showing that the Si/Ge layer serves as the glide mirror plane
for the glide operation ($M_z|\frac{1}{2}\frac{1}{2}0$).
}
\label{fig:lattice}
\end{figure}

\section{THEORY AND COMPUTATIONAL DETAILS}

As mentioned above, the ZrXY compounds 
crystallize in the tetragonal structure [see Fig. \ref{fig:lattice}(a)] with the \textit{P}4/\textit{nmm} 
(No. 129) nonsymmorphic space group which contains a glide mirror operation [Fig. \ref{fig:lattice}(b)]. 
The experimental lattice constants used in the present calculations are listed in Table I. 
The electronic band structures are calculated based on the DFT with 
the generalized gradient approximation (GGA) to the exchange-correlation potential \cite{perdew1996generalized}.
The accurate full-potential projecter-augmented wave (PAW) method\cite{blochl1994projector},
as implemented in the  Vienna \textit{ab initio} simulation package (VASP) \cite{kresse1996efficient,kresse1993ab},
is used. 
The valence electronic configurations of Zr, Ge, Te, Si, S, Se taken into account 
are $4s^24p^64d^25s^2$,  $3d^{10}4s^24p^2$, $5s^25p^4$, $3s^23p^2$, $3s^23p^4$, and $4s^24p^4$, respectively. 
A large plane wave cut-off energy of 400 eV is used. 
The self-consistent electronic band structure calculations are performed with a $\Gamma$-centered $k$-point mesh 
of 25 $\times$ 25 $\times$ 25 used in the  Brillouin zone integration by the tetrahedron method \cite{jepson1971,Temmerman89}.

The intrinsic SHC is calculated within the Berry phase formalism \cite{guo2008intrinsic,xiao2010berry}. 
In this approach, the SHC ($\sigma_{ij}^s=J_i^s/E_j$) is simply given by the Brillouin zone (BZ) integration of the spin Berry 
curvature for all occupied bands,
\begin{equation}
\label{eq:1}
\begin{aligned}
\sigma_{ij}^{s}=-e\sum_{n}\int_{BZ}\frac{d\textbf{k}}{(2\pi)^3}f_{\textbf{k}n}\Omega_{ij}^{n,s}(\textbf{k})
\end{aligned}
\end{equation}
\begin{equation}
\label{eq:2} 
\begin{aligned}
\Omega_{ij}^{n,s}=-\sum_{n'\neq n}\frac{2Im\left[\langle\textbf{k}n|\{\tau_{s},v_{i}\}/4|\textbf{k}n'\rangle\langle\textbf{k}n'|v_{j}|\textbf{k}n\rangle \right] }{(\epsilon_{\textbf{k}n}-\epsilon_{\textbf{k}n'})^2}
\end{aligned}
\end{equation}
where $J_i^s$ is the $i$th component of spin current density, $f_{\textbf{k}n}$ is the Fermi distribution, and $\Omega_{i,j}^{n,s} (\textbf{k})$ 
is the spin Berry curvature for the $n^{th}$ band at $\textbf{k}$. $i,j=x,y,z$ and $i\neq j$. 
$s$ is the spin direction, $\tau_s$ is the Pauli matrix, and $v_i$ is the velocity operator~\cite{Guo05}.
Similarly, the spin Nernst conductivity ($\alpha_{ij}^s=J_i^s/\Delta_jT$) can be written 
as\cite{xiao2010berry,guo2017large}
\begin{equation}
\label{eq:3} 
\begin{aligned}
\alpha_{ij}^{s}&= \frac{1}{T}\sum_{n}\int_{BZ}\frac{d\textbf{k}}{(2\pi)^3}\Omega_{i,j}^{n,s} (\textbf{k})\\
&\times [(\epsilon_{\textbf{k}n}-\mu)f_{\textbf{k}n}+k_{B}T\ln (1+e^{-\beta(\epsilon_{\textbf{k}n}-\mu)})].
\end{aligned}
\end{equation}
In the SHC and SNC calculations, the velocity $\langle\textbf{k}n'|v_{i}|\textbf{k}n\rangle$ 
and spin velocity $\langle\textbf{k}n|\{\tau_{s},v_{i}\}/4]|\textbf{k}n'\rangle$ matrix elements 
are calculated from the self-consistent relativistic band structures within the PAW formalism~\cite{Adolph01},
and the band summation and the BZ integration with the tetrahedron method are carried out 
by using a homemade program~\cite{Guo14,guo2017large}.
A very fine $k$-point mesh of 15498 k points in the IBZW is used, and  
this corresponds to the division of the $\Gamma X$ line into $n_d=40$ intervals. 
Further test calculatons for ZrSiS using denser $k$-point meshes of $n_d=50$ and 60 show 
that the calculated SHC and SNC converge within a few percent.

\section{RESULTS AND DISCUSSION}

\begin{table*}
\caption{Experimental lattice constants ($a$, $c$), calculated density of states at the Fermi level [$N(E_{F})$], 
spin Hall conductivity (SHC, $\sigma_{xy}^{z}$, $\sigma_{xz}^{y}$, $\sigma_{zx}^{y}$) and its energy derivative [$\sigma_{xy}^{z}(E_F)'$] 
at the Fermi level $(E_{F})$, as well as spin Nernst conductivity (SNC, $\alpha_{xy}^{z}$, $\alpha_{xz}^{y}$, $\alpha_{zx}^{y}$) at temperature
$T=300$ K. Previous results for noncollinear antiferromagnetic (AF) Mn$_{3}$Sn, ferromagnetic (FM) Mn$_{3}$Ga, nonmagnetic metal Pt and Weyl semimetal
TaAs are listed for comparison. Note that the unit of the SHC (SNC) is ($\hbar$/e)(S/cm) [($\hbar$/e)(A/m-K)].}
\begin{ruledtabular}
\begin{tabular}{c c c c c c c c c c c}
System &$a$&$c$&$N(E_{F}$) & $\sigma_{xy}^{z}$ & $\sigma_{xz}^{y}$ & $\sigma_{zx}^{y}$ &$\sigma_{xy}^{z}(E_F)'$ & $\alpha_{xy}^{z}$ & $\alpha_{xz}^{y}$ & $\alpha_{zx}^{y}$ \\
& (\AA ) & (\AA ) & (1/eV) &             &             &             & ($\hbar$/e)(S/cm-eV)&              &           &            \\ \hline
ZrSiS & 3.544$^a$ & 8.055$^a$& 0.75 & 79 & -280 & -611 & -1229 & 0.60 & 0.52 & 1.51 \\
ZrSiSe & 3.623$^a$ & 8.365$^a$ & 0.47 & -26.6 & -452 & -556 & -3143 & 0.40 & 0.61 & 1.02 \\
ZrSiTe & 3.692$^a$ & 9.499$^a$ & 0.96 & -197 & -376 & -755 & 1141 & -0.80 & -0.76 & -0.60\\
ZrGeS & 3.656$^a$ & 8.107$^a$ & 1.15 & 217 & -106 & -208 & -2624 & 0.39 & 0.29 & 0.54 \\
ZrGeSe & 3.706$^a$ & 8.271$^a$ & 0.76 & 182 & -247 & -486 & -1465 & 0.37 & 0.49 & 0.96 \\
ZrGeTe & 3.866$^a$ & 8.599$^a$ & 1.12 & 136 & -262 & -551 &-1064 & 0.53 & 0.48 & 1.17 \\
Mn$_{3}$Sn (AF)$^b$ & -- & -- & 1.96 & 72 & -- & -- & -845 & 0.91 & -- & -- \\
Mn$_{3}$Ga (FM)$^b$ & -- & -- & 6.82 & -678 & -- & -- & -10601 & 0.44 & -- & -- \\
fcc Pt & -- & -- & 1.75 & 2139$^c$ & -- & -- & 1214$^c$ & -1.09 (-0.91)$^b$,-1.57$^{d}$ & -- & -- \\
TaAs$^{e}$ & -- & -- & -- & -781 & -- & -- & -- & -- & -- & -- \\
\end{tabular}
\end{ruledtabular}
{$^{a}$X-ray diffraction experiment \cite{Haneveld64}; 
$^{b}$\textit{Ab initio} calculation \cite{guo2017large}; 
$^{c}$\textit{Ab initio} calculation \cite{guo2008intrinsic}; 
$^{d}$ Experiment at 255 K \cite{meyer2017observation}; 
$^{e}$\textit{Ab initio} calculation \cite{sun2016strong}.}
\label{table:1}
\end{table*}

\begin{table*}
\caption{\textit{P}4/\textit{nmm} symmetry-imposed shapes of the SHC and SNC tensors \cite{seeman2015}.}
\begin{tabular}{c c c c c c}
\hline\hline
& SHC & & & SNC & \\
\underline{$\sigma$}$^x$ & \underline{$\sigma$}$^y$ & \underline{$\sigma$}$^z$ & \underline{$\alpha$}$^x$ & \underline{$\alpha$}$^y$ & \underline{$\alpha$}$^z$ \\
\hline
\\
$\left(\begin{array}{ccc} 0 & 0 & 0\\0 & 0 & -\sigma_{xz}^{y}\\0 & -\sigma_{zx}^{y}&0\end{array}\right)$ & 
$\left(\begin{array}{ccc} 0 & 0 & \sigma_{xz}^{y}\\0 & 0 & 0\\\sigma_{zx}^{y} & 0&0\end{array}\right)$ &
$\left(\begin{array}{ccc} 0 & \sigma_{xy}^{z} & 0\\-\sigma_{xy}^{z} & 0 & 0\\0 & 0&0\end{array}\right)$ &
$\left(\begin{array}{ccc} 0 & 0 & 0\\0 & 0 & -\alpha_{xz}^{y}\\0 & -\alpha_{zx}^{y}&0\end{array}\right)$ & 
$\left(\begin{array}{ccc} 0 & 0 & \alpha_{xz}^{y}\\0 & 0 & 0\\\alpha_{zx}^{y} & 0&0\end{array}\right)$ &
$\left(\begin{array}{ccc} 0 & \alpha_{xy}^{z} & 0\\-\alpha_{xy}^{z} & 0 & 0\\0 & 0&0\end{array}\right)$ \\ \\
\hline\hline
\end{tabular}
\label{table:2}
\end{table*}

\subsection{Electronic band structures}
\begin{figure}[tbph] \centering
\includegraphics[width=8.6cm,height=10cm]{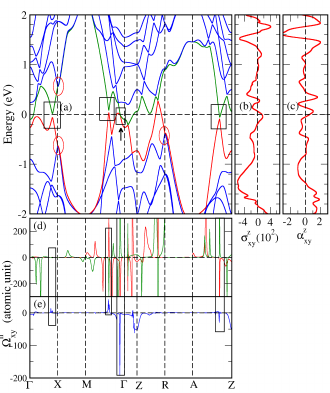}
\caption{ZrGeTe. (a) Relativisitc band strucuture, (b) spin Hall conductivity
($\sigma_{xy}^{z}$) as a function of energy, (c) spin Nernst conductivity ($\alpha_{xy}^{z}$)
at $T=300$ K as a function of energy, (d) band-decomposed spin Berry curvature $\Omega_{xy}^{n}$
and total spin Berry curvature $\Omega_{xy}$ along the high symmetry lines in the
Brillouin zone. In (a), the upwards black arrow indicates the gapped Type-II Dirac point.
In (a), (b) and (c), the Fermi level is at zero energy, and the unit
for the SHC (SNC) is ($\hbar$/e)(S/cm) [($\hbar$/e)(A/m-K)].
}
\label{fig:ZrGeTe}
\end{figure}

\begin{figure}[tbph] \centering
\includegraphics[width=8.6cm,height=10cm]{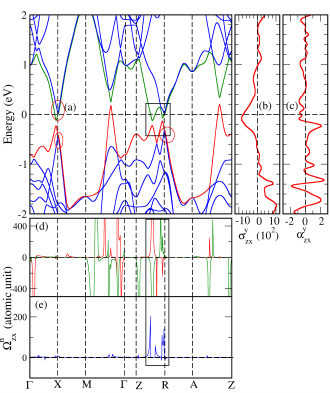}
\caption{ZrSiTe. (a) Relativisitc band strucuture, (b) spin Hall conductivity ($\sigma_{zx}^{y}$)
as a function of energy, (c) spin Nernst conductivity ($\alpha_{zx}^{y}$) at $T=300$ K as
a function of energy, (d) band-decomposed spin Berry curvature $\Omega_{zx}^{n}$
and total spin Berry curvature $\Omega_{zx}$ along the high symmetry lines in the
Brillouin zone. In (a), (b) and (c), the Fermi level is at zero energy, and the unit
for the SHC (SNC) is ($\hbar$/e)(S/cm) [($\hbar$/e)(A/m-K)].
}
\label{fig:ZrSiTe}
\end{figure}

The six compounds of the ZrSiS-family studied in this paper all have a non-magnetic ground state. 
The calculated relativistic band structures are shown in Fig. \ref{fig:ZrGeTe}(a) and Fig. \ref{fig:ZrSiTe}(a) 
for ZrGeTe and ZrSiTe, respectively, and that of the other compounds are displayed in 
Figs. S1 - S6 in the Supplemental Material (SM)\cite{supp}.
The bands near the Fermi level, labeled in green and red, are mainly contributed by Zr $d$-orbitals.
Their semimetallic nature can be seen from the calculated low density of states (DOS) near the Fermi level,
as listed in Table I. With the Kramers degeneracy given by the time-reversal symmetry together 
with the spatial inversion symmetry, all the compounds possess globally double degeneracy at every $k$-point. 
Furthermore, two doubly degenerate bands would cross at time-reversal invariant momenta (TRIM) X and R 
under the protection of the glide mirror symmetry [Fig. \ref{fig:ZrGeTe}(b)],
thus forming a four-fold Dirac node at these $k$-points. 
The Dirac nodes pinned at the TRIMs cannot be gapped by the SOC. 
The Dirac nodes at the X and R points are highlighted with the red circles in Figs. \ref{fig:ZrGeTe}(a) 
and Fig. \ref{fig:ZrSiTe}(a) as well as in Figs. S1-S6 in the SM \cite{supp}. 
These are just the end points of the Dirac line-nodes along the X-R and A-M lines 
at the zone boundaries\cite{Su2018,chen2017dirac}, which are protected by the glide mirror operation,
as labeled by the red line in e.g., Fig. \ref{fig:ZrGeTe-gxmg}(a).

Unlike the nonsymmorphic symmetry-protected line nodes, the unprotected nodes along the $\Gamma$-X, $\Gamma$-M, 
Z-R, and Z-A would be gapped out because the $C_{2v}$ symmetry has only one irreducible representation 
in the presence of the SOC. These gapped nodes can be seen from the calculated band structures 
in Fig. \ref{fig:ZrGeTe}(a) and Fig. \ref{fig:ZrSiTe}(a), where some of the gapped nodes are highligeted by
the black boxes with the corresponding spin Berry curvature peaks displayed in Fig. \ref{fig:ZrGeTe}(e) 
and Fig. \ref{fig:ZrSiTe}(e). Interestingly, the gapped Dirac nodes along these high symmetry lines 
form a diamond-shaped loop on the $\Gamma$-X-M plane and also the Z-R-A plane, as indicated by the green lines 
in Fig. \ref{fig:ZrGeTe-gxmg}(a). These nodal loop features in the band structures 
have been observed in the previous ARPES experiments\cite{schoop2016dirac,hosen2017tun,nakamura2019evidence}. 
Such diamond-shaped Fermi surface can also exhibit interesting properties such as 
large anisotropic magneto-resistance \cite{ali2016butterfly} and the U-shape optical 
conductivity \cite{allah2019optical}. In addition to the type-I Dirac nodes shared by the ZrSiS-family, 
ZrGeTe has interesting type-II Dirac points near the $\Gamma$ point just below the Fermi level, 
as indicated by the black arrow in Fig. \ref{fig:ZrGeTe}(a). These type-II Dirac points become slightly gapped 
(the gaps are smaller than 0.0001 eV) when the SOC is taken into account.

\subsection{Spin Hall effect}
The SHC ($\sigma_{ij}^{s}$; $s,i,j=x,y,z$) for a solid is a third-order tensor. 
However, for a highly symmetric crystal such as the ZrXY family, most of the tensor elements are zero. 
The nonmagnetic ZrXY family has space group \textit{P}4/\textit{nmm}, which corresponds to
the magnetic Laue group 4/\textit{mmm1'}. In Table II, the shape of the SHC tensor for the nonmagnetic ZrXY family,
uncovered by a symmetry analysis \cite{seeman2015}, is displayed.
Therefore, the nonmagnetic ZrXY family has only three nonzero elements, 
namely, $\sigma_{xz}^{y}$, $\sigma_{zx}^{y}$ and $\sigma_{xy}^{z}$. 
Note that the $x, y,$ and $z$ here denote the [100], [010], and [001] directions, respectively, 
and hence the $x$-direction is equivalent to the $y$-direction. 
Consequently, $\sigma_{yz}^{x} = -\sigma_{xz}^{y}$ and $\sigma_{zy}^{x} = -\sigma_{zx}^{y}$. 
In Table I, the calculated nonzero elements of the SHC of all six compounds are listed. 
The $\sigma_{xy}^{z}$ of some well studied systems 
such as noncollinear antiferromagnet Mn$_3$Sn and ferromagnetic Mn$_3$Ga\cite{guo2017large}, 
platinum metal \cite{guo2008intrinsic,meyer2017observation}, and 
Weyl semimetal TaAs \cite{sun2016strong}, are also listed in Table I for comparison. 

In Table I, the results for the six compounds considered here are listed in the increasing order of their SOC strengths. 
Among them, ZrSiS has the weakest SOC strength while ZrGeTe has the strongest one. 
This is based on the well-known fact that the larger the atomic number of an element is,
the stronger the SOC in the atoms of the element. The values of 
some nonzero elements of the SHC tensor in the six compounds can be large. In particular,
the $\sigma_{zx}^{y}$ of ZrSiTe has the largest SHC value of -755 ($\hbar$/e)(S/cm). 
This SHC value is ten times larger than that [72 ($\hbar$/e)(S/cm)] of the noncollinear 
antiferromagnetic Mn$_3$Sn \cite{guo2017large} and also slightly larger than the SHC [72 ($\hbar$/e)(S/cm)]
of the ferromagnetic Mn$_3$Ga [-678 ($\hbar$/e)(S/cm)] \cite{guo2017large}.
Interestingly, this is close to the calculated value of the $\sigma_{xy}^{z}$ of TaAs~\cite{sun2016strong},
where the spin Hall response was found to mainly come from the anticrossing Weyl nodal lines gapped by the SOC,
similar to the mechanism found here which will be illustrated below in the next subsection.
Nonetheless, Table I indicates that the values of all nonzero elements of the SHC tensor 
in the six compounds are smaller than the SHC ($\sigma_{xy}^{z}$) of fcc platinum, which possesses 
the largest intrinsic SHC [$\sim$2200 ($\hbar$/e)(S/cm)] among the transition metals \cite{guo2008intrinsic,Guo14}. 
However, we expect that the members of the ZrSiS family with a large SHC would 
have a higher charge-spin current conversion efficiency than platinum. 
The spin Hall angle ($\Theta_{sH}$), which is defined as the ratio
of the SHC to the ordinary electric conductivity, is a widely used measure of
this efficiency. Members of the ZrSiS family are semimetals and thus should
have an electric conductivity much smaller than that of platinum.
In particular, the in-plane conductivity of ZrSiTe at room temperature 
is about 10000 S/cm \cite{hosen2017tun}, which is six times smaller than that (64000 S/cm) of platinum\cite{Ando08}. 
Thus, we can expect that the $\Theta_{sH}$ for ZrSiTe would be twice as large as that of platinum.
Note that the conductivity along the $z$ axis (out-of-plane direction) in the ZrSiS family 
could be an order of magnitude smaller.~\cite{Novak19} 

Interestingly, all six compounds exhibit a strong anisotropy in the spin Hall effect.
In particular, Table I shows that the SHC of ZrSiSe could be increased by a factor of 20 
if the spin polarization of the spin current is switched from the $z$ direction (perpendicular to the layers)
to an in-plane direction (the $x$ or $y$ direction). Furthermore, in ZrGeY (Y = Te, Se, S), 
a rotation of the spin polarization direction from the $z$ axis to an in-plane direction
would change the sign of the SHC (see Table I).  
Table I also indicates that the SHC depends strongly on the direction of the applied eiectric field.
For example, the SHC of all the six compounds for the field applied along the [100] direction 
($\sigma_{zx}^{y}$) is nearly two times larger than that along the [001] direction ($\sigma_{xz}^{y}$) (Table I).
This large anisotropy in the SHE of the ZrSiS family could be attributed to their layered 
tetragonal structure with the four-fold rotational axis perpendicular to the layers (Fig. 1).  

Since the SHC could be sensitive to the location of the Fermi energy ($E_F$), 
we also calculate the SHC as a function of $E_F$. 
In panel (b) of Fig. \ref{fig:ZrGeTe} and Fig. \ref{fig:ZrSiTe}, 
the $\sigma_{xy}^{z}$ of ZrGeTe and $\sigma_{zx}^{y}$ 
of ZrSiTe are displayed as a function of $E_F$, respectively, and that for the other nonzero 
elements are presented in Figs. S1-S6 in the SM \cite{supp}. 
Indeed, as one can see from the figures, the SHC curves show a significant dependence
on the $E_F$. Interestingly, most of the SHC elements have a peak near the $E_F$
[see panel (b) in Figs. 2 and 3 as well as Figs. S1-S6 in the SM \cite{supp}].
As will be explained in the next subsection, this is due to the fact that
there are a lot of Dirac nodal points and lines near the $E_F$ which are slightly gapped
by the SOC in these Dirac semimetal compounds. 
In ZrSiTe, the prominent peak in the $\sigma_{zx}^{y}$ spectrum is located slightly below
the $E_F$ (-0.08 eV) [see Fig. 3(b)], and thus, a small lowering of the $E_F$
could increase the $\sigma_{zx}^{y}$ from -755 to -850 ($\hbar$/e)(S/cm). 
This could be accomplished by hole-doping of 0.07 e/f.u via chemical substitution or electric gating.

\subsection{Spin Berry curvature analysis}
To understand the origin of the large SHC in some of the six compounds, let us now take a look
at the band- and $k$-resolved spin Berry curvatures for the bands near the Fermi level.
According to Eq. (\ref{eq:1}), the SHC is simply given by the summation
of spin Berry curvatures on the occupied bands only. Interestingly, it has been shown that
when a band-crossing point such as a Dirac point becomes slightly gapped by the SOC,
the spin Berry curvatures usually appear as a pair of peaks with opposite signs, respectively,
on the upper and lower gapped bands in the vicinity of the $k$-point\cite{guo2008intrinsic}.
When both bands are occupied, the contributions to the SHC from these two peaks would cancel
each other. However, when only one band is occupied or the $E_F$ falls within the gap,
only one peak of spin Berry curvature would contribute to the SHC, thereby giving rise to a large SHC.
Therefore, we show the calculated spin Berry curvatures  $\Omega_{xy}^{z}$ and $\Omega_{zx}^{y}$ 
for the energy bands near the  $E_F$ in ZrGeTe and ZrSiTe in Fig. \ref{fig:ZrGeTe}(d) and
Fig. \ref{fig:ZrSiTe}(d), respectively. The total spin Berry curvatures which are a summation of
the spin Berry curvatures of all the occupied bands below the Fermi level at each $k$-point, are displayed
in Fig. \ref{fig:ZrGeTe}(e) and Fig. \ref{fig:ZrSiTe}(e). 
The calculated spin Berry curvatures for the other cases are displayed
in Figs. S1 - S12 in the SM \cite{supp}.

\begin{figure}[htbp] \centering
\includegraphics[width=8.6cm,height=4.2cm]{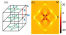}
\caption{(a) \textit{P}4/\textit{nmm} tetragonal Brillouin zone and topological nodes in ZrGeTe. Red lines denote
the nonsymmorphic symmetry-protected Dirac line nodes. Green curves indicate the unprotected
Dirac nodal lines which would become gapped in the presence of the SOC.
(b) Contour plot of the total spin Berry curvature $\Omega_{xy}({\bf k})$ of ZrGeTe
on the $\Gamma$-X-M plane. The large positive spin Berry curvatures can be seen on
the diamond-shaped nodal lines shown in (a), while the large negative spin Berry curvatures
on the four (red) points near the $\Gamma$ point come mainly from the gapped Type-II Dirac points
[see Fig. \ref{fig:ZrGeTe}(a)].}
\label{fig:ZrGeTe-gxmg}
\end{figure}

Figure \ref{fig:ZrGeTe}(a) shows that in the band structure of ZrGeTe, there are several gapped Dirac 
nodes near the $E_F$ along the $C_{2v}$ symmetry lines in the Brillouin zone, as highlighted by the black boxes. 
These gapped nodes result in prominent spin Berry curvature peaks shown in Fig. \ref{fig:ZrGeTe}(e).
Interestingly, the largest negative peak near the $\Gamma$ point comes from the SOC-gapped type-II Dirac point.
The relation between the large spin Berry curvatures and the Dirac nodes in the band structure can be seen more clearly
from Fig. \ref{fig:ZrGeTe-gxmg}. Figure \ref{fig:ZrGeTe-gxmg} (b) shows that the total spin 
Berry curvature $\Omega_{xy}({\bf k}$) positively peaks along the gapped nodal line loop in the $\Gamma$-X-M plane
[see the green diamond-shaped loop in Fig. \ref{fig:ZrGeTe-gxmg} (a)].
Four large negative spin Berry curvature peaks near the $\Gamma$ from the gapped type-II Dirac points 
[see Fig. \ref{fig:ZrGeTe}(a)] can also be seen in Fig. \ref{fig:ZrGeTe-gxmg} (b). 
Similar situations occur in the other five considered ZrSiS-type compounds (see Figs. S1-S12 in the SM\cite{supp}).
There are many more spin Berry curvature peaks along the diamond-shaped nodal lines 
than on the single nodal points. Therefore, the large SHC values found in
these Dirac semimetals originate mainly from the Dirac nodal lines.

Since there are many slightly gapped Dirac nodes near the Fermi level in the considered compounds
[see panel (a) in Figs. 2 and 3 as well as Figs. S1-S12 in the SM \cite{supp}],
one can imagine that a small variation of the Fermi level could significantly change
the SHC, as demonstrated in panel (b) of Figs. 2-3 and also of Figs. S1-S6 in the SM \cite{supp}.
Interestingly, this woud allow one to optimize the SHE in these Dirac semimetals by, e.g., chemical doping.
A surprisingly finding here is that the symmetry-protected nodes (degeneracies) can also generate large
spin Berry curvatures, which would form an antisymmetric butterfly-like shape on the two sides of the crossing point
[see, e.g., the inset in Figs. S7(a) and S8(a) in the SM \cite{supp}].
Here along the high $C_{4v}$-symmetry $\Gamma$-Z line, the band crossing
due to a band inversion is allowed. For example, Fig. S7(a) shows that in ZrGeTe,
a symmetry-protected crossing exists along the $\Gamma$-Z, due to the band inversion
between $\Gamma_{6-}, M_{6-}$ bands and $\Gamma_{7-},M_{7-}$ bands. The interesting shape of the peaks 
can be understood in terms of Eq. (\ref{eq:2}) as follows. When the order of $n$ and $n'$ is inverted on either side
of the crossing point, a sign change in the spin Berry curvature would occur, as can be see from Eq. (\ref{eq:2}).
Nonetheless, since these symmetry-protected Dirac nodes are well below the Fermi level, they
are not expected to contribute significantly to the SHC.

\subsection{Spin Nernst effect}
The SNC ($\alpha_{ij}^{s}$) for a solid is also a third-order tensor.
Since the transverse spin currents generated by a longitudinal electric field 
and a longitudinal temperature gradient have the same transformation properties under the symmetry operations,
the shape of the SNC is identical to that of the SHC \cite{seeman2015}, as displayed in Table II.  
Therefore, the SNC of the nonmagnetic ZrXY family also has only three nonzero elements 
of $\alpha_{xz}^{y}$, $\alpha_{zx}^{y}$, and $\alpha_{xy}^{z}$.
Table I shows that the calculated values of the nonzero elements of the SNC tensor of
the six compounds are significant. In fact, the $\alpha_{zx}^{y}$ of ZrGeTe and ZrSiS
are even larger than the $\alpha_{xy}^{z}$ of platinum (Table I).
Therefore, these findings suggest that the DSMs of the ZrSiS-family
can serve as promising candidate materials for spintronics
and spin caloritronics due to their efficient charge-spin current conversion driven by heat.

As for the SHC, the SNC of the six compounds is significantly anisotropic.
For example, Table I shows that the SNC of ZrGeTe for the spin polarization of the spin current along the $z$ axis
($\alpha_{xy}^{z}$) is about half of that for the spin polarization along the $y$ axis ($\alpha_{zx}^{y}$). 
Furthermore, the SNC of ZrGeTe for the electric field applied in the [100] direction ($\alpha_{zx}^{y}$)
is two times larger than that in the [001] direction ($\alpha_{xz}^{y}$).
Nevertheless, in contrast to the SHC, the SNC in ZrGeY (Y = Te, Se, S) does not change sign when the spin polarization
of the spin current is rotated from the $z$ axis to the $x$ or $y$  axis (Table I).

We also calculate the SNC as a function of $E_F$.
In panel (c) of Fig. \ref{fig:ZrGeTe} and Fig. \ref{fig:ZrSiTe},
the $\alpha_{xy}^{z}$ of ZrGeTe and $\alpha_{zx}^{y}$
of ZrSiTe are displayed as a function of $E_F$, respectively. For the other compounds,
the corresponding SNC curves are presented in Figs. S1-S6 in the SM \cite{supp}.
These figures suggest that the SNC has a stronger $E_F$ dependence than the SHC.
For instance, the $\alpha_{xy}^{z}$ of ZrGeTe increases rapidly as the $E_F$ increases up to $\sim$0.06 eV
where it reaches the local maximum of 1.0 ($\hbar$/e)(A/m-K). As the  $E_F$ further increases,
it decreases steadily and changes sign at $\sim$0.14 eV and then reaches the negative local
maximum of -0.93 ($\hbar$/e)(A/m-K) at 0.21 eV[see Fig. \ref{fig:ZrGeTe}(c) and also Fig. S1(c)]. 
Remarkably, as the $E_F$ rises, the $\alpha_{zx}^{y}$ of ZrSiS decreases steeply, changes sign at 0.05 eV,
and then reaches the negative local maximum of 1.99 ($\hbar$/e)(A/m-K) at 0.10 eV, 
just slightly above the Fermi level [see Fig. S6(c) in \cite{supp}].
A similar behavior is found for the $\alpha_{zx}^{y}$ of ZrSiSe [Fig. S5(c) in \cite{supp}].
On the other hand, the $\alpha_{zx}^{y}$ of ZrSiTe would change sign and then become as large as 
2.15 ($\hbar$/e)(A/m-K) when the $E_F$ is lowered to -0.23 eV [see Figs. 3(c) ánd 5(b) as well as Fig. S4(c) in \cite{supp}]. 
Again, these interesting phenomena could be observed in these compounds by chemical doping or electric gating.

\begin{figure}[tbph] \centering
\includegraphics[width=8.6cm,height=5.8cm]{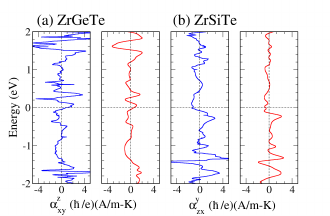}
\caption{Verification of the Mott relation [Eq. (\ref{eq:4})]. The SNC at $T=300$ K
as a function of the Fermi level, calculated from the energy derivative of the SHC
using Eq. (\ref{eq:4}) (blue line) and also from the relativistic band structure
using Eq. (\ref{eq:3}) (red line) for (a) ZrGeTe and (b) ZrSiSe.}
\label{fig:Mott}
\end{figure}

In order to understand the origin of the prominent features in the spectrum of the SNC
as a function of $E_F$, we note that in the low temperature limit, Eq. (\ref{eq:3}) 
can be written as the Mott relation,
\begin{equation}
\label{eq:4} 
\begin{aligned}
\alpha_{xy}^{z}(E_F)=-\frac{\pi^2}{3}\frac{k_B^2T}{e}\sigma_{xy}^{z}(E_F)',
\end{aligned}
\end{equation}
which indicates that the SNC is proportional to the energy derivative of the SHC 
at the $E_F$. In other words, a peak in the SNC would occur when the SHC
has a steep slope. In Table I, the energy derivatives of the $\sigma_{xy}^{z}$ at the $E_F$ 
for the six compounds are listed, and they agree in sign 
with the $\alpha_{xy}^{z}$ [note the minus sign in Eq. (\ref{eq:4})]. 
In Fig. \ref{fig:Mott}(a) [\ref{fig:Mott}(b)], the $\alpha_{xy}^{z}$ [$\alpha_{zx}^{y}$] estimated 
using the calculated $\sigma_{xy}^{z}(E)'$ [$\sigma_{zx}^{y}(E)'$] 
and the Mott relation [Eq. (\ref{eq:4})] (the blue curves) for ZrGeTe [ZrSiTe] is displayed together 
with the SNC calculated from the relativistic band structure using Eq. (\ref{eq:3}) (the red curves)
(See Fig. S13 in the SM \cite{supp} for the other four compounds).
Figure \ref{fig:Mott} and Fig. S13 show that most of the prominent peaks in the SNC are reproduced
by the Mott relation, suggesting that these peaks stem from the large energy-derivatives of the corresponding SHC.
Note that the rapid oscillations of the SNC derived from the Mott relation, e.g., 
around 2.0 eV in Fig. \ref{fig:Mott}, are due to the uncertainties in the numerical differentiation 
of the small oscillations in the SHC. 
Consequently, the plateau feature in the SNC between $E_F$ and -1.0 eV in ZrGeSe, ZrGeS, ZrSiSe, 
and ZrSiS can be understood as a result of the rather non-dispersive SHC within the same energy range, 
where few gapped Dirac points exist. Similar to the SHC, the SNC values for these six compounds
can be optimized by slightly tuning the Fermi level via, e.g., chemical substitution.
\\

\begin{figure}[tbph] \centering
\includegraphics[width=8.6cm,height=6cm]{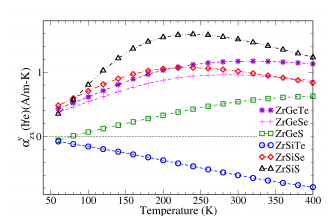}
\caption{Spin Nernst conductivity ($\alpha_{zx}^{y}$) as a function of temperature for the six considered compounds. }
\label{fig:T-dep}
\end{figure}

We also calculate the SNC as a function of temperature ($T$). The $T$-dependences of the $\alpha_{zx}^{y}$ 
for the six considered compounds are shown in Fig. \ref{fig:T-dep} and that of the other nonzero elements are presented
in Fig. S14 in the SM \cite{supp}. 
Figure \ref{fig:T-dep} shows that ZrSiS has the largest $\alpha_{zx}^{y}$ for $T \ge 100$ K.
The $\alpha_{zx}^{y}$ of ZrSiS reaches the maximum of $\sim$1.6 [($\hbar /e$)(A/mK)] 
at $T=240$ K and then decreases monotonically as $T$ further increases. 
Interestingly, the $\alpha_{zx}^{y}$ values of ZrGeTe, ZrGeSe, and ZrSiSe are very close
and have a very similar behavior, i.e., they all increase monotonically 
as $T$ goes from 60 K, and then converge to similar values up to $T=400$ K (Fig. \ref{fig:T-dep}).
In contrast, the $\alpha_{zx}^{y}$ of ZrGeS is negative at $T=60$ K and then changes sign 
at $T=80$ K. As $T$ further increases, it increases monotonically up to $\sim$400 K.
We notice that the $\alpha_{xz}^{y}$ of ZrGeS
exhibit an almost identical behavior to that of $\alpha_{zx}^{y}$ [see Fig. 6 and Figs. S14(b) in the SM \cite{supp}].
The $\alpha_{zx}^{y}$, $\alpha_{xz}^{y}$ and $\alpha_{xy}^{z}$ of ZrSiTe are negative in the entire considered temperature range
and decrease monotonically with temperature [see Fig. 6 and Figs. S14(b) in the SM \cite{supp}]. 
We note that the behavior of the other
two nonzero SNC elements ($\alpha_{xz}^{z}$ and $\alpha_{xz}^{y}$) of ZrSiTe are very similar 
(see Figs. S14 in the SM \cite{supp}). 

\section{CONCLUSIONS}
In summary, we have thoroughly studied the SHE and SNE in Dirac nodal-line semimetals 
ZrXY (X = Si, Ge; Y = S, Se, Te) by performing systematic \textit{ab initio} calculations 
within the Berry phase formalism. The calculated SHC and SNC of the six considered materials 
are large. In particular, the calculated SHC of ZrSiTe is as large as -755 [($\hbar / e$)(S/cm),
being smaller but in the same order of magnitude of that of platinum, which posseses the
largest intrinsic SHC among the metals. However, since the ZrSiS semimetals have a much smaller
electric conductivity, they are expected to have a larger spin Hall angle than platinum.
The calculated SNC of ZrSiS is $\alpha_{zx}^{y}$= 1.51 ($\hbar / e$)(A/mK), being even 
larger than that of platinum. More importantly, it is found that both the amplitude and sign of 
the SHC and SNC of these compounds
can be significantly manipulated by tuning the direction of the applied electric field or
spin current, and also can be optimized by slightly varying the Fermi level via, e.g., chemical doping.  
This indicates that these materials can be useful materials for spintronics and spin caloritronics.
Furthermore, the calculated band- and $k$-resolved spin Berry curvatures
reveal that the large SHE and SNE in these compounds and their remarkable tunabilities originate 
from the large number of Dirac nodal points and lines in the vicinity of the Fermi level.
In the presence of the SOC, these Dirac nodes are slightly gapped, resulting in large spin Berry curvatures,
and thus make major contributions to the SHC and SNC. 
This work is thus expected to stimulate further experimental investigations
such as spin-related transport measurements on these interesting materials.

\section*{ACKNOWLEDGMENTS}
The authors acknowledge the support from the Ministry of Science and Technology, 
the Ministry of Education, the National Center for Theoretical Sciences, and the Academia Sinica
of the R. O. C. The authors also thank Dr. Tien-Ming Chuang 
for valuable discussion on the electronic structures of the ZrSiS-type compounds
and Mr. Tzu-Cheng Wang for helpful communications on the electronic structure calculation of ZrSiS.



\end{document}